\documentclass[pra,twocolumn]{revtex4}
\usepackage[dvips]{graphicx}%
\usepackage{amsmath,amssymb,bm,color}
\newcommand{\braket}[2]{\langle #1 | #2 \rangle}
\newcommand{\ketbra}[2]{| #1  \rangle   \langle #2  |}
\newcommand{\ket}[1]{\left |  #1 \right \rangle}
\newcommand{\bra}[1]{ \left \langle #1  \right |}
\newcommand{\ave}[1]{  \langle #1  \rangle}
\def \tr{{\textrm {Tr}}}

\begin{document}
%%%%%%%%
%  \small
%%%%%%%%
\title{Einstein-Podolsky-Rosen-like correlation on a coherent-state basis \\ and inseparability of two-mode Gaussian states}
\author{Ryo Namiki}\affiliation{Department of Physics, Graduate School of Science, Kyoto University, Kyoto 606-8502, Japan}

\date{December 6, 2012}% \today}%December 6, 2012}%
\begin{abstract} 
The strange property of the Einstein-Podolsky-Rosen (EPR) correlation between two remote
physical systems is a primitive object on the study of quantum entanglement. In order to understand
the entanglement in canonical continuous-variable systems, a pair of the EPR-like uncertainties is
an essential tool. Here, we consider a normalized pair of the EPR-like uncertainties and  introduce a
state-overlap to a classically correlated mixture of coherent states. 
 The separable
condition associated with this state-overlap determines the strength of the EPR-like correlation on
a coherent-state basis in order that the state is entangled.  We show that the coherent-state-based condition is capable of detecting the class of two-mode Gaussian entangled states. We also present an experimental measurement scheme for estimation of the state-overlap by a heterodyne measurement and a photon detection with a feedforward operation.

 \end{abstract}

%\pacs{03.67.Dd, 42.50.Lc} 
%\keywords{quantum cryptography}
\maketitle

\section{Introduction}
In the seminal paper \cite{EPR35}, Einstein, Podolsky, and Rosen (EPR) considered a pair of particles, say $A$ and $B$, that possesses perfect correlation not only in their positions $x_{A (B)} $ but also in their momentums $p_{A (B)} $. From such a correlation, one can predict either the position or the momentum of one particle with certainty by measuring the position or the  momentum of the other particle, and this seemingly contradicts the canonical uncertainty relation 
\begin{eqnarray}
\sqrt{\ave{ \Delta^2 \hat x} \ave{\Delta^2 \hat p} } \ge |[\hat x,\hat p ]|/2=:C  , \label{ccr0}
\end{eqnarray} where %the deviation operator  is defined by 
 $ \Delta \hat O := \hat O - \ave{ \hat O } $.
 This type of seeming inconsistency between the quantum correlation and the canonical uncertainty relation is often termed as the EPR paradox and  has been providing insightful aspects on foundations of quantum physics and theory of entanglement \cite{Horo09,Ades07,gtphysrep, EPR-para}.

The EPR-type correlation is normally described by the variances of the EPR-type operators $\hat x_A- \hat x_B$ and $\hat p_A+\hat p_B$, and   the measured uncertainties can be a signature of quantum entanglement.  Duan \textit{et al.},  \cite{Duan00} have introduced the EPR-like operators $\hat X := |a| \hat x_A- \frac{1}{a}\hat x_B$ and $\hat P ' := |a| \hat p_A+ \frac{1}{a}\hat p_B$ with a real number $a$, and presented an inseparable condition associated with the total variance of the operators: A bipartite state is entangled if it violates the inequality 
\begin{eqnarray}
% \ave{ \Delta^2 \left(|a| x_A- \frac{1}{a}x_B \right)} +\ave{\Delta^2\left( |a| p_A+ \frac{1}{a}p_B\right) } %  \nonumber \\ 
\ave{ \Delta^2\hat X} +\ave{\Delta^2 \hat P' }  
\ge  2(a^2+ \frac{1}{a^2})  C  \label{D-sum-c}. 
\end{eqnarray}  
Interestingly,  this condition is conducted to determine the inseparability of two-mode Gaussian states. To be specific, for  any given entangled two-mode Gaussian state,  there exists a proper local Gaussian-unitary transformation and a parameter $a$ so that the inequality of Eq. (\ref{D-sum-c}) is violated. 
Its implication is that the origin of the inseparability of two-mode Gaussian states lies on the strength of the EPR-like correlation.

A quantum state on a bipartite system $AB$ is called separable if its density operator can be written in the convex sum form of the products of local density operators as $\rho_{AB} = \sum_i p_{i} (\rho_i )_A \otimes (\sigma_i )_B $ where $(\rho_i )_A$ and $(\sigma_i )_B$ are local density operators of the system $A$ and $B$, respectively, and $p_{i}$ is a probability distribution that satisfies $p_i \ge 0$ and $\sum p_{i} =1 $.
 A quantum state is said to be  
 entangled if it is  not separable.
  The separable density operator preserves its positivity under the transpose of its local density matrix. This property of positive partial transposition cannot hold for many of entangled density operators,   and non-positivity of the partial transposition is a signal of entanglement \cite{Peres}. An important fact is that the class of Gaussian entangled states belongs to the  entanglement with non-positive partial transposition \cite{Simon00,Giedke01}. It is shown that many of known separable conditions concerning the continuous-variable states, which include Eq. (\ref{D-sum-c}), can be derived by using partial transposition for moments of the annihilation operators and the creation operators \cite{SV,Mira09}.

 In quantum optics, the canonical variables correspond to the phase-space quadratures of optical modes, and their statistics can be measured by the homodyne measurement. This enables us to determine the moments of annihilation and creation operators in experiments.  The homodyne measurement is a standard Gaussian-measurement and  plays a central role in the continuous-variable quantum information processing \cite{CV-RMP}.  Another important Gaussian measurement is the heterodyne (double homodyne) measurement.  It measures the complex amplitude $\alpha$ of an optical coherent state $\ket{\alpha}$ and
  gives the projection probability to the coherent state $\bra{\alpha } \rho\ket{\alpha} $.
  The canonical quadratures and coherent-state amplitudes provide similar   phase-sensitive information of the optical modes, and both of them are thought to be useful to observe the properties of Gaussian states.    There have been several approaches to suggest the relation between heterodyne statistics and entanglement mainly associated with the transmission of coherent states \cite{GroInf}. It might be also insightful to consider the separablity problems related to  the phase-space distribution \cite{Bana99,Mar09,Mira10}. However, their implication with respect to    the EPR-like correlation has little been discussed.

    Recently, a separable condition with the state-overlap to the Gaussian distributed phase-conjugate pairs of coherent states was derived \cite{namiki11}. It states that  %. It shows that 
 any separable state has to satisfy
   \begin{eqnarray}
\left\langle \int p_\lambda (\alpha) \ket{ \alpha  }\bra{ \alpha  } \otimes \ket{\sqrt  \eta  \alpha ^*  }\bra{\sqrt  \eta \alpha ^* }  d^2\alpha  \right\rangle \le \frac{\lambda}{ 1+ \lambda+ \eta },  \label{overlapori} % \nonumber
\end{eqnarray}  where $p_\lambda( \alpha ) :=  \frac{\lambda }{\pi} \exp (- \lambda |\alpha |^2 ) $ and $\eta, \lambda  \ge 0 $. % and the probability distribution is given by \begin{eqnarray}p_\lambda( \alpha ) :=  \frac{\lambda }{\pi} \exp (- \lambda |\alpha |^2 )  \label{prior} . \end{eqnarray}
 Since the state-overlap in the left-hand side is written in terms of the projections to the coherent states, it can be estimated by  the statistics of the heterodyne measurement. The condition of Eq. (\ref{overlapori})  was formulated for the quantum benchmark problem \cite{Bra00,Ham05,namiki07,namiki08,Namiki11APB,namiki11,Has08,Has09,Cal09,Owari08,Adesso08,Guta10,Has10,Takano08,Kil10,Kil11,Namiki12R}, however, its utility and  significance  for the separability problem have little been discussed.

In this paper we investigate the properties of the overlap condition of Eq. (\ref{overlapori}) for the separability problem. We argue that the state-overlap is a form of the EPR-like correlation in a coherent-state basis. It is shown that the separable condition with the overlap and the separable condition with the EPR-like uncertainties can be formulated in parallel, and the violation of the separable conditions can be interpreted as %essentially the same 
a phenomenon to infer  the EPR paradox. 
For the Gaussian states given in a standard form of the covariance matrix we find a simple embrace relation between the separable conditions. This relation provides a geometric proof that the overlap condition can be conducted to determine the inseparability of two-mode Gaussian states. We also consider experimental measurement schemes to detect the state-overlap.

This  paper is organized as follows. We introduce  a limitation of the phase-space localization as a sort of the canonical uncertainty relation in Sec. II. We derive the separable condition with the EPR-like uncertainties and investigate its properties in Sec. III.  We derive the overlap separable condition and discuss its properties in Sec. IV. Then, we   address the utility of the overlap condition for Gaussian states in Sec. V. We present the experimental scheme   in Sec. VI. We conclude this paper in Sec. VII. % I.  

\section{Uncertainty relation and Phase-space Localization}
\begin{figure}[htbp]
\includegraphics[width=\linewidth]{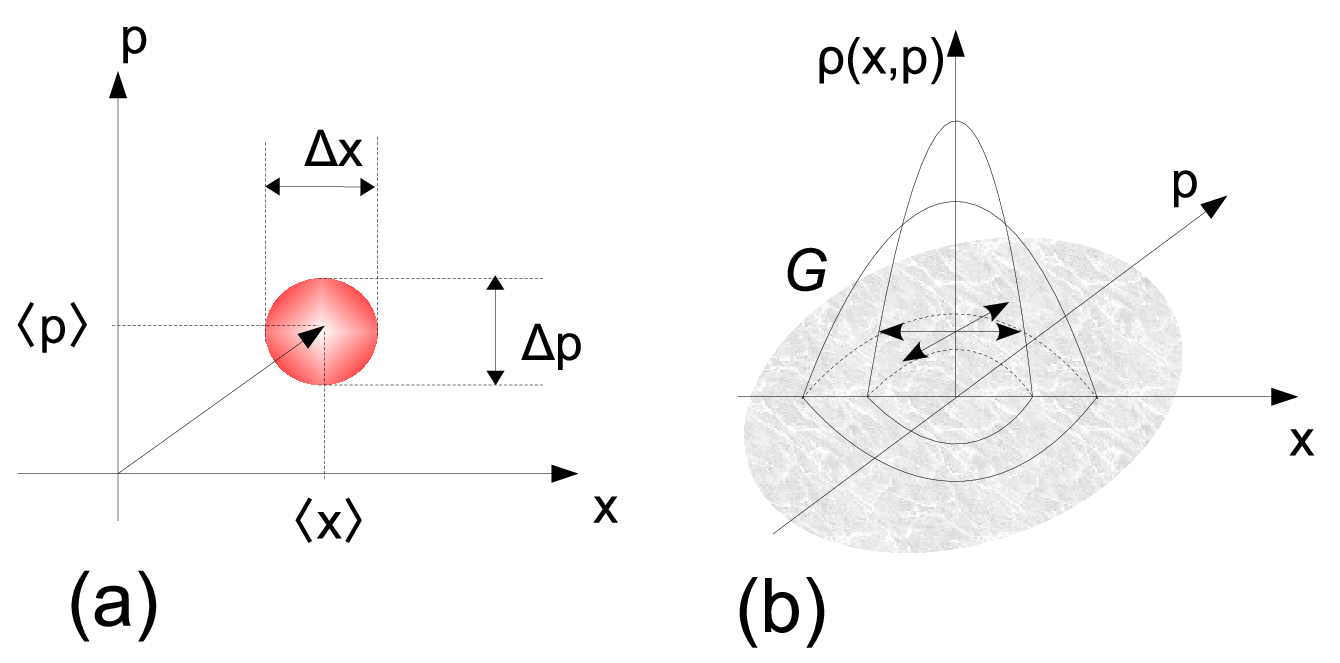}
  \caption{(Color online) The uncertainty relation gives a limitation on the localization in the phase space. (a) A physical state is spread in the phase space so that its volume of  $ {\Delta x} {\Delta  p} $ is no smaller than the minimum uncertainty product of  $|[\hat x, \hat p]|/2$ due to the canonical uncertainty relation. (b)   Another measure of the phase-space localization can be given by the convolution between the state distribution $\rho (x,p) $ and a localized distribution function $ G$ on the phase space. The value of the convolution designates    the concentration of the state distribution at the peak of the function $G$.   }  \label{fig:epp-fig1.eps}
\end{figure}

An intuitive interpretation of the canonical uncertainty relation in Eq. (\ref{ccr0}) is that the quantum state is located on the phase space with a finite volume [See FIG. \ref{fig:epp-fig1.eps}(a)]. When the volume is measured in terms of the uncertainty product $ \sqrt{\ave{\Delta^2 \hat x}\ave{\Delta^2 \hat p }}$, it cannot be smaller than the limit determined by the canonical commutation relation, i.e., $\sqrt{\ave{\Delta^2  \hat x}\ave{\Delta ^2  \hat p }} \ge |[\hat x, \hat p]|/2 $. 
 The standard deviation describes typical width of the phase-space distribution and thus the uncertainty product indicates a degree of localization of the phase-space distribution. Here, we consider another measure of the phase-space localization and present another form of the physical limitation.

Let us consider the density operator of a thermal state 
\begin{eqnarray}
\hat G_\lambda &: =& %\frac{\lambda}{ \pi} \int e^{-\lambda | \alpha | ^2} 
\int p_\lambda (\alpha  ) \ket{\alpha} \bra{\alpha} d^2 \alpha \nonumber \\
& =& \frac{\lambda}{1+\lambda } \sum_{n= 0}^{ \infty} \left(\frac{1 }{1+\lambda }\right) ^n \ket{n}\bra{n}. \label{eq-1}
\end{eqnarray}
Here we use the standard notation for the number state $\ket{n}$ and the coherent state $\ket{\alpha}= \sum _{n=0} ^\infty e^{-| \alpha |^2 /2 }  \alpha  ^n \ket{n} /\sqrt {n !}$. The phase-space distribution of the thermal state is an isotropic Gaussian distribution and 
 peaked at the origin of the phase space $\alpha = 0 $ [See FIG. \ref{fig:epp-fig1.eps}(b)]. 
 The  expectation value of the thermal state $\ave{\hat G }_\rho :=\tr {\hat G \rho}=\lambda \int Q_\rho (\alpha )e^{-\lambda |\alpha |^2} d^2\alpha $ is a Gaussian convolution of the Husimi-$Q$ function $Q_\rho (\alpha ):= \bra{\alpha}\rho \ket{\alpha}/\pi$. It suggests how strong the probability distribution is concentrated  %, whose size is typically determined by $\lambda^{-1}$,
 around the origin. Hence, it is likely that the expectation value is maximized by the state which has a sharply peaked $Q$ function at $\alpha =0 $. However, we cannot make the width of the $Q$ function arbitrarily small, and thus $\ave{\hat G_\lambda }$ has an upper bound. This upper bound offers another form of the physical limitation on the degree of the  phase-space localization.    From the second line of Eq. (\ref{eq-1}), an upper bound of $\ave{\hat G} $ is given by the maximum eigenvalue of the thermal state $\| \hat G_\lambda \|$  as  \begin{eqnarray}
\ave{\hat G_\lambda} \le   \| \hat G_\lambda \|= \frac{\lambda}{1+\lambda }. \label{eq2}
\end{eqnarray}
 %$\| \hat G_\lambda \|$ stands for the maximum eigenvalue of $G$ and this limit 
 This relation serves as a sort of the uncertainty relation, namely,  one cannot localize the physical state on the phase space so that the expectation value $\ave{\hat G} $ surpasses the physical limit $\| \hat G_\lambda \|$. We refer to $\ave{\hat G_\lambda }_\rho = \tr  \hat G_\lambda \rho $ as the $\lambda$-\textit{localization} of a density operator $\rho$. The equality of Eq. (\ref{eq2}) can be achieved by the vacuum state $\ket{0}$ and  the vacuum state is the maximally $ \lambda$-localized state. Since, the $\lambda$-localization is an overlap between a given state and the thermal state, it represents the probability of finding the states in the thermal distribution.

In order to see an intuitive connection between  the uncertainty product and the $\lambda$-localization, let us consider the case where the $Q$ function has a single peak at the origin. Let  $\delta x$ and  $\delta p$ be the width of the $Q$ function along the real $x$  direction and the imaginary $p$ direction, respectively. Then, the normalization condition  $1= \int Q_\rho (\alpha ) d^2 \alpha \simeq Q_\rho (0) \delta  x \delta  p$ implies $ Q_\rho (0) \simeq  (\delta  x \delta  p )^{-1}$. Hence, for sufficiently large $\lambda$, we have $\ave{\hat G_\lambda}=\lambda \int Q_\rho (\alpha )e^{-\lambda |\alpha |^2} d^2\alpha   \simeq \pi Q_\rho (0) \simeq  \pi  (\delta  x \delta  p )^{-1}$, namely, the $\lambda$-localization is  proportional to the inverse of the uncertainty product, in a certain limit.

\section{Separable conditions with the EPR-like uncertainties} 
 In this section, we derive a separable condition with a normalized EPR-like uncertainty product using partial transposition for the canonical uncertainty relation. This separable condition is called the product condition \cite{Reid89,Tan99,Gio03,Hyl06} and has a simple embrace relation to the sum separable condition of Eq. (\ref{D-sum-c}). In contrast to the sum condition, any point of the separable boundary of the product condition can be achieved by the product of the squeezed states. It is shown that the maximum of the EPR-like correlation can be achieved by a two-mode squeezed state (TMSS).  

We start with the canonical uncertainty  relation \begin{eqnarray}
\ave{\Delta^2 \hat x }\ave{\Delta^2 \hat  p } \ge \left( \frac{|[\hat x, \hat p]|}{2} \right ) ^2 =   C ^2.  \label{CCR}
\end{eqnarray}
 By introducing an ancilla system $B$ and applying a beamsplitter transformation $(\hat x_A, \hat  p_A) \to ( u \hat x_A - v  \hat x_B ,  u  \hat p_A - v  \hat p_B )$ we have \begin{eqnarray}
%& &\ave{\Delta^2  ( R^\dagger  \hat x R ) }\ave{\Delta^2 ( R^\dagger  \hat  p  R ) } \nonumber\\
%&= &
\ave{\Delta^2  \left( u \hat x_A - v  \hat x_B \right) }\ave{\Delta^2 \left( u  \hat p_A - v  \hat p_B \right)} \ge  C  ^2, \label{normalur}%\left( \frac{|[\hat x, \hat p]|}{2} \right ) ^2 
\end{eqnarray} where we assign the index $A$ for the original system and assume that the real parameters $(u,v)$ satisfy %are real and the parameter  %action of the beamsplitter transformation is given by 
%\begin{eqnarray}(\hat x_A, \hat  p_A) \to ( u \hat x_A - v  \hat x_B ,  u  \hat p_A - v  \hat p_B )\label{BStrans}\end{eqnarray}
% are real parameters satisfying
 the relation  $u^2+v^2=1$. % and $u>0$. %, $u>0$.
 When we make the replacement  
$p_B \to - p_B$ \cite{Aga05} we have a product separable condition  \cite{Gio03,Hyl06,SV}
 \begin{eqnarray}
\ave{\Delta^2  \left( u \hat x_A - v  \hat x_B \right) }\ave{\Delta^2 \left(   u \hat p_A + v \hat p_B \right)} \ge  C ^2  %\left( \frac{|[\hat x, \hat p]|}{2  } \right ) ^2
 .  \label{prod-c}
\end{eqnarray}
The replacement corresponds to the transposition of the system $B$ with respect to the number basis (see the below proof). The left-hand side of Eq. (\ref{prod-c}) is a normalized EPR-like uncertainty product so that it becomes a normal uncertainty product for the canonical variables under the partial transposition as in Eq. (\ref{normalur}). Since the partial transposition is not a physical transformation, it is no reason to consider that Eq. (\ref{prod-c}) holds for all physical states.
 We can show that separable states cannot violate this inequality as follows: 
 
  \textit{Proof. --- }
  Let us write $\hat X := u  \hat x_A - v  \hat x_B $, $\hat P := u \hat p_A - v  \hat p_B $, and the partial transposition, which transposes the system $B$ with respect to the number basis,  
\begin{eqnarray}
\Gamma : (|l \rangle  \langle m | \otimes|j \rangle  \langle k |  ) \to (|l \rangle  \langle m | \otimes|k \rangle  \langle j | ). \label{defPTG}
\end{eqnarray}
 For product states, we can write the expectation value of the partial transposed observable $\hat O$  as $\ave{\Gamma [ \hat O ] }_{\phi \otimes \varphi }= \tr[\Gamma [ \hat O ] \ket{\phi}\bra{\phi}\otimes \ket{\varphi}\bra{\varphi} ]=  \tr[  \hat O    \Gamma(\ket{\phi}\bra{\phi}\otimes \ket{\varphi}\bra{\varphi} ) ]  =\tr( \hat O    \ket{\phi}\bra{\phi}\otimes \ket{\varphi ^* }\bra{\varphi ^*} )=   \ave{\hat O}_{\phi \otimes \varphi^* }$. Here we defined the conjugate state by $\ket{\varphi ^*}:= \sum_{n=0}^\infty \braket{\varphi} {n} \ket{n}$. Since  the off-diagonal elements  of the position operator $\hat x = \sqrt{ C }( \hat a +\hat a^\dagger ) $ in  the number basis are real we have $\Gamma [ \hat X ] =\hat X $. In contrast,  the  off-diagonal elements  of the  momentum operator $\hat p = \sqrt{ C }( \hat a -\hat a^\dagger )/i$ in the number basis are pure imaginary, and  we thus have $\Gamma [ \hat P ] = u \hat p_A + v  \hat p_B $. Noting that $\Gamma [ \hat P^2 ] =(\Gamma [ \hat P ])^2 $ and $\Gamma [ \hat X^2 ] =(\Gamma [ \hat X ])^2 $, we can estimate the left-hand side of Eq. (\ref{prod-c}) as   $\ave{\Delta^2 (\Gamma \hat X)}_{\phi \otimes \varphi } \ave{\Delta^2( \Gamma \hat P) }_{\phi \otimes \varphi }    = \ave{\Delta^2 \hat  X}_{\phi \otimes \varphi^* }\ave{\Delta^2 \hat  P}_{\phi \otimes \varphi^* } \ge \min_{\phi \otimes \varphi }\left\{ \ave{\Delta^2 \hat  X}_{\phi \otimes \varphi } \ave{\Delta^2 \hat P}_{\phi \otimes \varphi } \right\} \ge (|[\hat X, \hat P]|/2)^2 $ for any product state.
    Hence, for any separable state  $ \rho_s = \sum_i p_i \ket{\phi_i}\bra{\phi_i}\otimes \ket{\varphi_i}\bra{\varphi_i}$ we have 
\begin{eqnarray}
&&\ave{\Delta^2 (\Gamma \hat X)}_{ \rho_s }\ave{\Delta^2 (\Gamma \hat P)}_{ \rho_s }\nonumber\\ &=& (\sum_i p_i \ave{\Delta^2 \hat  X}_{\phi_i \otimes \varphi_i^* }) (\sum_j p_j \ave{\Delta^2 \hat  P}_{\phi_j \otimes \varphi_j^* } )\nonumber \\
&\ge&  \left(\sum_j p_j \sqrt{\ave{\Delta^2 \hat X}_{\phi_j \otimes \varphi_j^* }  \ave{\Delta^2 \hat  P}_{\phi_j \otimes \varphi_j^* }} \right)^2 \nonumber\\
&\ge & (|[\hat X, \hat P]|/2)^2=  C ^2. 
\end{eqnarray}
 From the second line to the third line, we set $a_j= \sqrt{ p_j \ave{\Delta^2 \hat X}_{\phi_j \otimes \varphi_j^* }  } $, $b_j= \sqrt{ p_j \ave{\Delta^2 \hat  P}_{\phi_j \otimes \varphi_j^* }  } $ and use the Schwarz inequality $|\Vec a|^2 |\Vec b|^2 \ge |\Vec a \cdot \Vec b |^2$.  \hfill   $\blacksquare$ 

  An important implication of the product separable condition of Eq. (\ref{prod-c}) is that the EPR-like correlation cannot be stronger than the canonical uncertainty limit without entanglement. As was shown in the proof, the EPR-like uncertainty product for a product state can be mapped into the canonical uncertainty product for its conjugat state when the EPR-like operators are normalized so that their partial transpositions form a pair of the canonical observables. From this normalization,  the phase-space localization  can be directly associated with the EPR-paradox, namely, the seeming violation of the limitation on the phase-space localization verifies the existence of entanglement.

\begin{figure}[htbp]
\includegraphics[width= 0.8\linewidth]{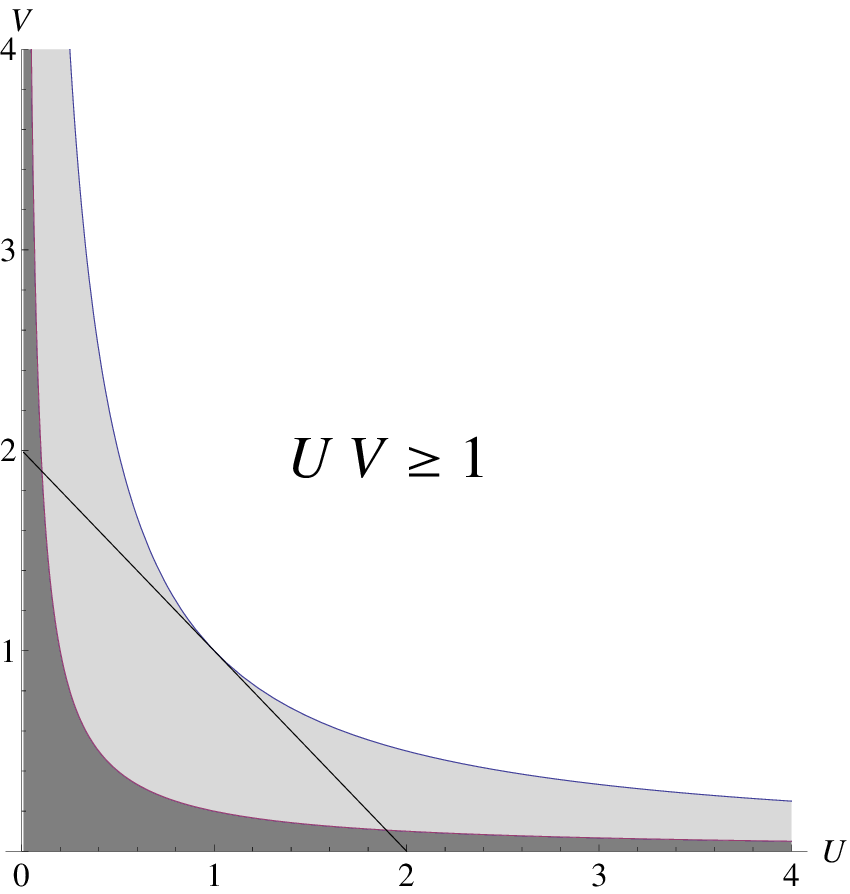}%{epp-1101-c1.eps} %,epp-ur-fig1.eps}%{epp-fig-c.eps}
  \caption{(Color online)  Relation between the inseparability and the normalized EPR-like uncertainties $U$ and $V$ in Eq. (\ref{defUV}).  The product condition of Eq. (\ref{prod-c}) [Eq. (\ref{uv1})] implies that any state is entangled if its location $(U,V)$ is below the inverse proportional curve $UV = 1$ (gray regime). On the other hand, for any point of this curve one can find a corresponding separable state (See main text).  The violation of the sum condition of Eq. (\ref{sum-c}) [Eq. (\ref{uv2})] occurs for the states located below the  straight line  $U + V =2 $, and the entangled states detected by the sum condition belong to a subset of the entangled states detected by the product condition. The dark gray regime is physically unaccessible due to the global canonical uncertainty relation of Eq. (\ref{p-boundary}). In this figure we set $|u^2-v^2|^2= 0.2 $ specifically so that the physical boundary is given by $UV=  0.2$.   \label{fig:epp-fig1-1.eps}}
\end{figure}

   Dividing both sides of Eq. (\ref{D-sum-c}) by $(a^2 +1/a^2)$ we have a normalized sum condition 
\begin{eqnarray}%\ave{\hat D}:= 
\ave{\Delta^2  \left( u \hat x_A - v  \hat x_B \right) }+ \ave{\Delta^2 \left( u  \hat p_A + v  \hat p_B \right)}\ge 2  C,  %|[\hat x, \hat p]|
 \label{sum-c}
\end{eqnarray} where we set 
\begin{eqnarray}
(u,v )= \frac{1}{\sqrt{a^2 +a^{-2}}}( |a| ,  a^{-1} ). \label{a-u}
\end{eqnarray}
  This sum condition of Eq. (\ref{sum-c})
 can be also obtained by taking the square root on both sides of Eq. (\ref{prod-c}) and using the relation $\ave{\Delta^2  \left( u \hat x_A - v  \hat x_B \right) }+ \ave{\Delta^2 \left( u  \hat p_A + v  \hat p_B \right)} \ge 2 \sqrt{\ave{\Delta^2  \left( u \hat x_A - v  \hat x_B \right) }  \ave{\Delta^2 \left( u  \hat p_A + v  \hat p_B \right)}}$.
 From this derivation we can see that the inequality of Eq. (\ref{prod-c}) is  automatically violated if the inequality of Eq. (\ref{sum-c}) is violated. This suggests that the product condition of Eq. (\ref{prod-c}) is better to detect entanglement than the sum condition of Eq. (\ref{sum-c}). 
To show the advantage of the product condition more clearly, let us write the normalized uncertainties of  the EPR-like operators as  
\begin{eqnarray}
  U: &=&    {  \ave{\Delta^2  \left( u \hat x_A - v  \hat x_B \right) }} /C  \nonumber \\ V:&= & { \ave{\Delta^2 \left( u  \hat p_A + v  \hat p_B \right)}}/C . \label{defUV}
\end{eqnarray}
%, and set the nomarization of the commutator $[\hat x, \hat p]= i$.
 Then,         
the product separable condition of Eq. (\ref{prod-c}) leads to 
\begin{eqnarray}
  U  V  \ge 1, \label{uv1}
\end{eqnarray}
and the sum separable condition of Eq. (\ref{sum-c}) leads to
\begin{eqnarray}
  U  +  V \ge 2.  \label{uv2}
\end{eqnarray}
The embrace relation between Eqs. (\ref{uv1}) and (\ref{uv2}) can be directly observed in FIG. \ref{fig:epp-fig1-1.eps}.  
Since there is no separable state below the curve $UV=1$ (gray regime of FIG. \ref{fig:epp-fig1-1.eps}), 
 the states located on the boundary of Eq. (\ref{sum-c}) should be entangled except for the single point %that satisfies %$\ave{\Delta^2  \left( u \hat x_A - v  \hat x_B \right) }= \ave{\Delta^2 \left( u  \hat p_A + v  \hat p_B \right)}$
 $(U,V)=(1,1)$. % in FIG. . 
As a result, the sum condition fails to notice the entangled states located in the area enclosed by the two curves $UV=1 $ and $ U +V =2 $.  
As a separable state located on the boundary of Eq. (\ref{uv1}),  we can find the product of the squeezed states $\hat S_A\otimes \hat S_B | 0,0 \rangle_{AB} $ where $\hat S$ stands for the squeezing operator.

 Note that the physical limitation for the EPR-like uncertainty product is give by 
 \begin{eqnarray}
& & \ave{\Delta^2  \left( u \hat x_A - v  \hat x_B \right) }\ave{\Delta^2 \left(   u \hat p_A + v \hat p_B \right)} \nonumber \\
 &\ge&  \left( \frac{ | [ u \hat x_A - v  \hat x_B ,   u \hat p_A + v \hat p_B ]|}{ 2} \right)^2  =  (u^2-v^2)^2  C ^2    %\left( \frac{|[\hat x, \hat p]|}{2  } \right ) ^2
 . \nonumber % \label{prod-c}
\end{eqnarray} In terms of $U$ and $V$, it can be expressed as 
\begin{eqnarray}
  U  V  \ge |u^2-v^2|^2 .\label{p-boundary}
\end{eqnarray}   
This inequality is saturated by the TMSS
\begin{eqnarray}
| \psi_\zeta \rangle_{AB} &=& \sqrt{1-| \zeta | ^2} \sum_{n= 0}^\infty\zeta ^n    |n \rangle  _A |  n \rangle_B \label{def-TMSS}, 
\end{eqnarray}
with $\zeta =v/u <1 $. We can observe that the TMSS is located at $(U,V)= (  {u^2-v^2},   {u^2-v^2})$ on the $U$-$V$ plane and that the physical boundary of Eq. (\ref{p-boundary}) can be covered by the state $\hat S_A\otimes \hat S_B | \psi_\zeta \rangle_{AB} $ similar to the case that the product of the squeezed states covers the separable boundary of Eq. (\ref{uv1}). 
Noting the relation $U+V \ge 2 \sqrt{UV}\ge 2 ({u^2- v^2})$, we can see that the physically possible minimum of the sum $U+V$ is also achieved by the same TMSS  located at $(  {u^2-v^2},   {u^2-v^2})$.
%the point of 

  The fact that the product condition is better than the sum condition is generally stressed in  \cite{Hyl06,Gio03}  and the results of Refs.  \cite{Hyl06,SV,Gio03} essentially include the condition of Eq. (\ref{prod-c}) although the EPR-like operators are not normalized so that their partial transpositions form a pair of the canonical variables.   For the case of $|u|=|v|$, the separable condition of  Eq. (\ref{prod-c}) is derived in \cite{Tan99,Manc02,Aga05}. 
 From the superiority of the product condition and the fact \cite{Duan00,Simon00} that any Gaussian entangled state can be detected 
by the violation of the sum condition, it is concluded \cite{Gio03} that any two-mode Gaussian entangled state can be detected by the violation of the product condition. There is an approach to consider that the sum condition is a condition for a quadratic Hamiltonian, thereby a separable condition on the variance of the normalized Hamiltonian is derived \cite{namiki10}.

 To estimate the left-hand sides of Eqs. (\ref{prod-c}) and (\ref{sum-c}) in the experiments, one may perform the joint quadrature measurement of $\hat x_A \hat x_B$ and $\hat p_A  \hat  p_B$ (For the estimation of the covariance matrix of the two-mode state, it requires the measurement of $\hat x_A \hat p_B$ and $\hat p_A \hat x_B$, additionally). The measured homodyne statistics of $\hat x_A \hat x_B$ and $\hat p_A  \hat  p_B$  give the six variances $\{ \ave{\Delta^2 \hat x_A }$, $\ave{\Delta^2 \hat x_B}$, $\ave{\Delta^2 \hat p_A }$, $\ave{\Delta^2 \hat p_B}$, $\ave{\Delta  \hat x_A \Delta \hat x_B}$, $\ave{ \Delta \hat p_A \Delta \hat  p_B}\}$. Then, the left-hand side of  Eq. (\ref{prod-c}) can be determined for any   set of $(u,v)$.  In practice, it is better to use the set of the parameters $(u, v)$ so that the left-hand side of Eq. (\ref{prod-c}) becomes as small as possible. The minimum can be readily found by setting $(u, v) = (\cos  {\theta}  , \sin  {\theta}) $ and plotting the left-hand side of  Eq. (\ref{prod-c}) as a function of $\theta$. The parameter $\theta$ corresponds to the effect of the  global rotation  by the beamsplitter transformation. Although the joint  squeezing $\hat S_A \otimes\hat S_B$ belongs to the set of local operations, it is not easy to access experimentally. In turn, to achieve the boundary of the product condition, the sum condition requires additional local squeezing operations. This suggests actual experimental advantage to use the product condition in place of the sum condition.

 Note that the left-hand side of Eq. (\ref{sum-c}) becomes a quadratic form of $\Vec u := (u,v)^t$ as  $   \Vec u^t  (M_x+ M_p)  \Vec u  $ 
where 
\begin{eqnarray}
M_x:&=&  \left(
  \begin{array}{cc}
        \ave{\Delta^2 \hat x_A } &   - \ave{\Delta \hat x_A \Delta \hat x_B }  \\
    -  \ave{\Delta \hat x_A \Delta \hat x_B }   &   \ave{\Delta^2 \hat x_B }  \\
  \end{array}
\right) \nonumber \\
M_p:&=&  \left(
  \begin{array}{cc}
        \ave{\Delta^2 \hat p_A } &   \ave{\Delta \hat p_A \Delta \hat p_B }  \\
      \ave{\Delta \hat p_A \Delta \hat p_B }   &   \ave{\Delta^2 \hat p_B }  \\
  \end{array}
\right). \label{mat1}
\end{eqnarray}
Hence, the minimum of the left-hand side of Eq. (\ref{sum-c}) is given by  the minimum eigenvalue of the matrix $M_x+M_y$. The minimum plays an important role in   Refs. \cite{Mar09,Fijikawa09}. By using the matrices of Eq. (\ref{mat1}), the left-hand sides  of Eq. (\ref{prod-c}) can be expressed in a compact form
$ ( \Vec u^t  M_x \Vec u ) ( \Vec u^t  M_p \Vec u )  $. %the analysis of

\section{separable condition with the coherent-state-based EPR-like correlation}In this section we use the partial transposition for the limitation on the $\lambda$-localization of Eq. (\ref{eq2}), and derive the overlap separable condition corresponding to Eq. (\ref{overlapori}). It determines  the strength of the EPR-like correlation in a coherent-state basis in order that the  state is entangled. The maximal correlation on this basis is also obtained by a TMSS.

Let us consider the following positive operator 
\begin{eqnarray}
{\hat G_\lambda '} :&=& \hat R \hat G_\lambda \otimes \ket{0}\bra{0} \hat  R^\dagger\nonumber \\
 &=& \int p_\lambda (\alpha) %\frac{\lambda }{  \pi} \int e^{- {\lambda} | \alpha | ^2}
  \ket{v \alpha} \bra{ v \alpha}  \otimes \ket{ {u }   \alpha} \bra{ {u }   \alpha}  d^2 \alpha,   \label{16} %\nonumber \\ %& =& \frac{1}{1+\lambda } \sum_{n= 0}^{ \infty} \left(\frac{1 }{1+\lambda }\right) ^n \ket{n}\bra{n}. 
\end{eqnarray}
where the thermal state $\hat G _\lambda$ is defined in Eq. (\ref{eq-1})  and  the beamsplitter transformation $\hat R$ is defined through its action on the coherent state $\hat R \ket{\alpha }\ket{0}= \ket{v \alpha }\ket{u \alpha} $. 
 Since the spectrum of $ \hat G_\lambda'$ is the same as the spectrum of $\hat G_\lambda$, the physical limitation for the $\lambda$-localization  of Eq. (\ref{eq2})  also holds for $ \hat G_\lambda'$ as \begin{eqnarray}
\ave{\hat G_\lambda ' }  %&=& \frac{\lambda^{(2)}}{ \pi} \int e^{-\lambda^{(2)}  | \alpha | ^2} \ave{ \ket{\alpha} \bra{\alpha}  \otimes \ket{\alpha} \bra{\alpha} } d^2 \alpha  \nonumber \\
 & \le & \frac{\lambda }{1+\lambda } %= \lambda_0 
 . \label{LG2}
\end{eqnarray}
The equality is achieved by the product of the vacuum states $ \ket{0,0}_{AB}$. %, and $\ket{0}\ket{0}$ is the most localized state. % of $\lambda$.

From the partial transposition of $\hat G'$ and the physical limitation of Eq. (\ref{LG2})  we obtain the overlap separable condition \cite{namiki11}: 
\begin{eqnarray}
\ave{ \Gamma \hat G_\lambda ' }  %&=& \ave{\frac{\lambda }{ \pi} \int e^{- \lambda   | \alpha | ^2} \ket{v \alpha} \bra{v \alpha}  \otimes \ket{u \alpha^*} \bra{u \alpha^*}  d^2 \alpha } \nonumber \\  
& \le & \frac{\lambda }{1+\lambda }, \label{LGTP2}
\end{eqnarray} where  the partial transposition can be written as %of  $\hat G' $
\begin{eqnarray}\Gamma(\hat G_\lambda')= %\frac{\lambda }{ \pi} \int e^{- \lambda   | \alpha | ^2}
 \int p_\lambda (\alpha)  \ket{v \alpha} \bra{v \alpha}  \otimes \ket{u \alpha^*} \bra{u \alpha^*}  d^2 \alpha .  \label{GamG}\end{eqnarray}
Here, the action of the partial transposition map $\Gamma$ of Eq. (\ref{defPTG}) induces the complex conjugation of the coherent-state amplitude of the second system in Eq. (\ref{16}). The equality of Eq. (\ref{LGTP2}) is also achieved by the product of the vacuum states $ \ket{0,0}_{AB}$. 
We can show that the overlap condition of Eq. (\ref{LGTP2}) holds for any separable state as follows:

\textit{Proof.--- } For any separable state $\rho_s$, $\Gamma [\rho_s] $ is a density operator. Using Eq. (\ref{LG2}) for  $\Gamma [\rho_s] $, we have   $\ave{ \Gamma \hat G'  }_{\rho_s} = \tr ( \rho_s  \Gamma [ \hat G' ] )= \tr ( \Gamma [ \rho_s]   \hat G'  )\le \lambda/(1+\lambda ) $. Hence, the violation of the condition in Eq. (\ref{LGTP2}) implies that the state is entangled. \hfill $\blacksquare$

\begin{figure}[htbp]
\includegraphics[width=.8\linewidth]{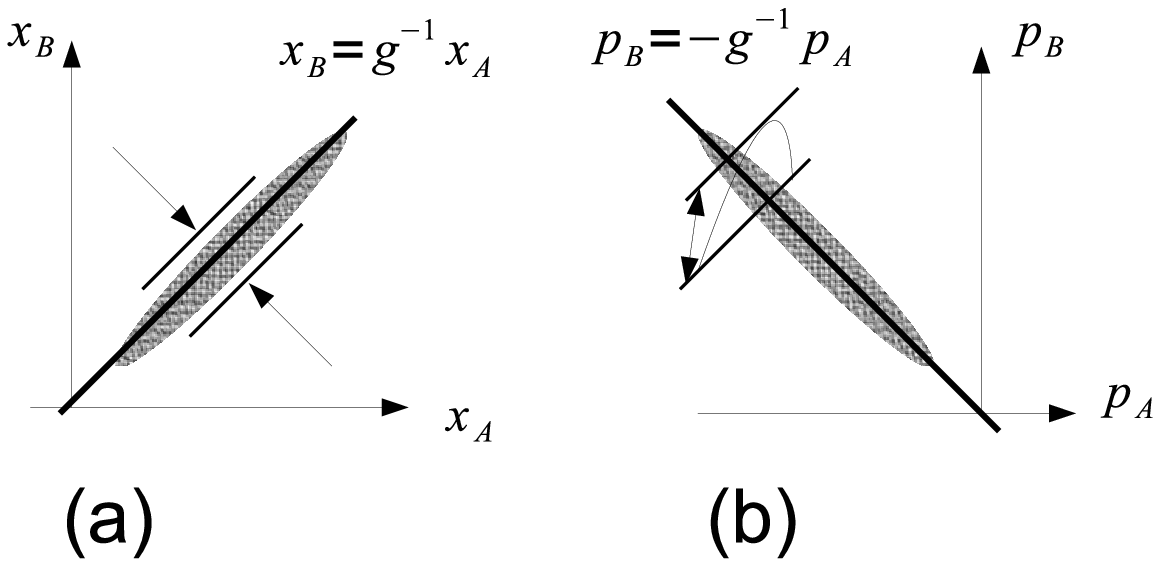}
 \caption{A pair of the EPR-like particles exhibits a strong positive correlation on their positions $x_A$ and $x_B$. It also exhibits a strong negative correlation on their momentums $p_A$ and $p_B$. A total deviation from the lines of $x_B= g^{-1}x_A$ and $p_B= - g^{-1}p_A$ represents the strength of the EPR-like correlation. The deviation can be directly related to the width of the correlation distribution  as (a). The 
 strength of the correlation can also be related to the intensity of the distribution  as (b). 
The two conditions of $x_B= g^{-1}x_A$ and $p_B= - g^{-1}p_A$ are combined into the single expression $\alpha_B^*= g^{-1} \alpha_A $ with the complex amplitudes $\alpha_A = x_A+ ip_A $ and  $\alpha_B = x_B+ ip_B $. This suggests another form of the EPR-like correlation in terms of the distribution intensities associated with  the  pairs of coherent states  $\{\ket{\alpha}_A \ket{g\alpha^*}_B \}$.   }  \label{fig:epp-fig2.eps}
\end{figure}

 The expectation value of $\Gamma \hat G _\lambda '$ in Eq. (\ref{GamG}) is a weighted sum of the probability that the pair of coherent states $\ket{\alpha}_A\ket{g \alpha ^*}_B$ is contained in the given state, where $g=u/v$ is a real number. Recalling that the complex amplitude is defined as $\alpha = x +i p $, the state-overlap $\ave{\Gamma \hat G_\lambda'}$ essentially represents the strength of the  EPR-like correlation so that the relations $x_A = g x_B$ and $p_A =-g p_B$  hold, simultaneously. Actually the two relations can be combined to the single expression $\alpha_A = g \alpha _B^* $ (See also FIG. \ref{fig:epp-fig2.eps}).  Note that we can reproduce  Eq. (\ref{overlapori}) from Eq. (\ref{LGTP2}) as follows. We change the variable of integration $\alpha '= v \alpha $ and 
 make the replacement $ \lambda /v^2 \to \lambda $ on  Eq. (\ref{LGTP2}). Then, we can obtain  Eq. (\ref{overlapori})  by setting   $\eta = \frac{u^2}{v^2} = \frac{1}{v^2} -1 $. An expression of $\ave{\Gamma \hat G _\lambda '}$ for Gaussian states  and its link to the separable conditions with the EPR-like uncertainties can be found in Sec. V.

A violation of the overlap separable condition of Eq. (\ref{LGTP2}) can be observed for the TMSS. From Eq. (\ref{def-TMSS}) and  Eq. (\ref{GamG}) we have 
\begin{eqnarray}
\ave{\Gamma (\hat G_\lambda' )}_{\psi_\zeta} =\frac{\lambda ( 1-\zeta ^2 )}{u^2 [1+ \lambda/ u^2- \zeta ^2 +( v/u - \zeta)^2  ]} . \label{15lhs}
\end{eqnarray} %where the TMSS is given by 
If we set $\sqrt 2u = \sqrt 2v  =1$ and $2 \zeta = 1$, the right-hand side of Eq. (\ref{15lhs}) becomes $3 \lambda  /(2+4 \lambda)$. Hence, for $\lambda <1 $, we can observe $\ave{\Gamma (\hat G_\lambda' )}_{\psi_\zeta} >\lambda /(1+\lambda )$. If the transformation $\Gamma$ preserved the phase-space localization, this expression would imply the violation of the physical limitation  for the $\lambda$-localization of Eq. (\ref{LG2}) as a sort of the EPR paradox. In reality, the partial transposition  $\Gamma$ is not a physical transformation  and it is no need to consider that the violation of Eq. (\ref{LGTP2}) violates the physical limitation of Eq. (\ref{LG2}). 
The paradox, in which entangled states can be ``localized'' beyond the limit achieved by the pair of coherent states, is essentially identical to the  phenomenon that the EPR-like uncertainty  product violates the canonical uncertainty limit.  %discussed in the previous section.
 It might be helpful to consider that the phenomenon comes from the use of a strange way to sum the phase-space volume based on the partial transposed unit of the volume measure, in which the sign of the local momentum is inverted $p_B \to - p_B$. This inversion suggests the complex conjugation $i \to - i $ because the sign of the commutation relation is changed due to the replacement $[\hat x, \hat p] \to - [\hat x, \hat p]$. Such a replacement affects the coherence between the two  systems and some of entangled states exhibit seemingly abnormal phase-space volume. 

Note that $\Gamma \hat G_\lambda ' $ of Eq. (\ref{GamG}),  as a density operator,  is located at the point $(U,V) = (1,1) $ on the separable boundary of the $U$-$V$ plane as the vacuum state $\ket{0}\ket{0}$ is located at the same point (See FIG. \ref{fig:epp-fig1-1.eps}).  Moreover, the state obtained by applying the collective local squeezing both on A and B to $\Gamma \hat G_\lambda ' $,  i.e.,  $\hat S_A \otimes \hat S_B  \Gamma (\hat G_\lambda ' ) \hat S_A^\dagger \otimes \hat S_B^\dagger  $  moves along the local minimum uncertainty boundary of Eq. (\ref{uv1}) as $\hat S_A \otimes \hat S_B \ket{0,0}_{AB}$ does. We can see that $\hat S_A \otimes \hat S_B  \Gamma ( \hat G_\lambda '  ) \hat S_A^\dagger \otimes \hat S_B^\dagger  $ reduces to the product of the pure squeezed states $\hat S_A \otimes \hat S_B \ket{0,0}_{AB}$  in the pure limit $\lambda \to \infty $. Although the form of the mixture shows the correlation explicitly, its EPR-like correlation is no stronger than the correlation given by the uncorrelated state $\ket{0,0}$. % when it is measured  either by the state-overlap in Eq. (\ref{LGTP2})  or by the product of the uncertainties in Eq. (\ref{prod-c}). product

 As was mentioned above,  the strength of the coherent-state-based EPR-like correlation $\ave{\Gamma \hat G_\lambda '}$ is  a state-overlap to the classically correlated state. It simply suggests the probability that the state contains the conjugate coherent-state pairs, and the separable condition of Eq. (\ref{LGTP2}) gives the threshold  of the  pair appearance in order that the state is  entangled.  
The maximum of the coherent-state-based EPR-like correlation $\ave{\Gamma (\hat G_\lambda' )}$ is given by the operator norm of $\Gamma (\hat G_\lambda' )$ as \begin{eqnarray}
\max_\rho \ave{\Gamma (\hat G_\lambda' )}_\rho  &=& 
 \|\Gamma (\hat G_\lambda' ) \| \nonumber \\ &=&\frac{4}{(\nu_+ +1)(\nu_- +1)} \nonumber \\
&=& \frac{2 \lambda}{1+ \lambda + \sqrt{(1+ \lambda )^2 -4 u^2v^2 }}
\end{eqnarray}
where we use  the symplectic eigenvalues $\nu_\pm$ of $\Gamma (\hat G_\lambda' )$ defined in Eq. (\ref{SyEi}).  
This maximum value is achieved by the TMSS of Eq. (\ref{def-TMSS}) when we set 
\begin{eqnarray}
\zeta= [(1+ \lambda)- \sqrt{(1+ \lambda)^2 -4 u^2 v^2}] /(uv). 
\end{eqnarray}
Hence, the EPR-like correlation can be maximized by the TMSS on the coherent-state basis as well as on the basis of the quadrature uncertainties. 
 
 In general, it is not necessary to choose the symmetric Gaussian distribution to discuss the localization. We can proceed similar discussion with a wide class of  distributions. For any positive operator with a positive-$P$ representation  $\rho_{pp}$ on a single mode,  a two-mode operator $\hat   R \rho_{pp}\otimes \ket{0}\bra{0} \hat  R ^\dagger $ is an unnormalized separable state and the following relation holds 
$\ave{\hat R \rho_{pp} \otimes \ket{0}\bra{0} \hat  R^\dagger}\le \| \rho_{pp}  \|$  since %\begin{eqnarray}
$\ave{ \rho_{pp} } \le %\max_{\braket{\phi}{\phi}=1 }=  \bra{\phi}  \rho_{pp} \ket{\phi}=:
\| \rho_{pp}  \| $. 
%\end{eqnarray}
By taking the partial transposition we have a separable condition
\begin{eqnarray}
\ave{\Gamma [\hat R  \rho_{pp}  \otimes \ket{0}\bra{0} \hat  R^\dagger] }\le \|\rho_{pp}  \|.  \label{eq22} \end{eqnarray} If we know the  $P$ representation  $ \rho_{pp}= \int P(\alpha ) \ketbra{\alpha }{\alpha } d ^2 \alpha $, the partial transposition can be calculated as $\Gamma [\hat  R  \rho_{pp}  \otimes \ket{0}\bra{0} \hat  R^\dagger]=\int P(\alpha )\ket{v\alpha}\bra{v\alpha} \otimes \ket{u\alpha^*}\bra{u\alpha^*} d^2\alpha $.  
The condition of Eq. (\ref{eq22})  with a non-Gaussian distribution of $P(\alpha )$ might be useful when the expectation value $\ave{ \ketbra{u\alpha }{u\alpha } \otimes  \ketbra{v\alpha }{v\alpha } }$ is obtained for a limited number of the amplitude $\alpha $ in the real experiments. In such a case, one can choose $P (\alpha )$ as a discrete distribution  associated with the observed set of the amplitudes.   
Further, analysis of potential utilities of this approach beyond the case of the symmetric Gaussian distribution is left for future works.

\section{EPR-like correlation for the detection of two-mode Gaussian entanglement}
In this section we apply the overlap condition of Eq.  (\ref{LGTP2}) for the two-mode Gaussian states in a standard form of the covariance matrix. In the flat-distribution limit ($\lambda \to 0$), the overlap condition can be described by  the normalized EPR-like uncertainties similar to the cases of the sum condition and the product condition in Sec. III. We find a simple embrace relation on these separable conditions. This relation geometrically proves that 
 the coherent-state-based approach is capable of detecting the inseparability of  all two-mode Gaussian  states.

Let us consider the covariance matrix of a two-mode state $\rho$ 
\begin{eqnarray}
\gamma_\rho := \langle \Delta \hat d  \Delta \hat d^t + ( \Delta \hat d   \Delta \hat d^t)^t \rangle _{  \rho}  %- 2 \langle \hat R \rangle _{ \rho}   \langle \hat R^t  \rangle _{ \rho} 
% \nonumber 
\end{eqnarray} 
where $\hat d := (\hat x_A,\hat p_A,\hat x_B, \hat p_B)^t $.  The physical requirement for the covariance matrix is given by $\gamma +i \Omega \ge 0$ where  
\begin{eqnarray}
\Omega: = \left(
  \begin{array}{cc}
       J &  0   \\
       0   &  J\\
      \end{array} 
\right) = J \oplus J, \ \ 
 J := \left(
  \begin{array}{cc}
       0 &  1   \\
       -1   &  0\\
      \end{array}  \label{defJ1}
\right)
\end{eqnarray}
 with % $ I_2 = \textrm{diag} (1,1)$  and
  the normalization  $[\hat x_A, \hat p_A]= [\hat x_B, \hat p_B]= i $ (We set $C=|[\hat x, \hat p ]|/2 =1/ 2 $ henceforth).  The characteristic function of a two-mode density operator $\rho$ is defined by
\begin{eqnarray}
\chi (\xi) := \tr [\rho \exp (i \hat d^t  \xi)] \end{eqnarray}
where $\xi = (\xi_1,\xi_2,\xi_3,\xi_4)^t  \in \mathbb R^4 $
is a real vector. The density operator $\rho$ can be written by the inverse of the Fourier transform as
\begin{eqnarray}
\rho =  {(2\pi ) ^{-2}} \int_{\mathbb R ^4} \chi (\xi) \exp (- i \hat d^t  \xi)d   \xi . \label{defIFT}
\end{eqnarray}
 We call the state is a Gaussian state if its  characteristic function is Gaussian as 
\begin{eqnarray}
 \chi (\xi)=\exp (i d^t \xi - \frac{1}{4} \xi^t \gamma \xi ), \label{defGC}
\end{eqnarray}
where $d = \ave{\hat d} = \tr[ \rho \hat d] $ is the mean of the phase-space position. The two-mode Gaussian state is completely characterized by its covariance matrix $\gamma$ and the mean $d$.    
The mean $d$ can be freely chosen by applying local displacement operators, and is irrelevant to the inseparability. Hence, we consider the zero-mean case $d=0$ in  the following discussion. 
  
The operator $\Gamma( \hat G_ \lambda ') $ of Eq. (\ref{GamG}) is a density operator of a Gaussian state and its covariance matrix is given by 
\begin{eqnarray}
\gamma_0 =I_4 +  \frac{2}{\lambda}\left(
  \begin{array}{cc}
     v^2 I_2 &  uv Z   \\
     uv Z   & u^2   I_2   \\
      \end{array}  \label{bd1}
\right),  \end{eqnarray} where $ I_4 = \textrm{diag} (1,1,1,1)$, $ I_2 = \textrm{diag} (1,1)$, and $Z= \textrm{diag}(1,-1 )$. The symplectic eigenvalues \cite{Ades07,namiki11R,comS} of $\gamma _0$ is given by 
\begin{eqnarray}
\nu_\pm = \sqrt{(1+\lambda)^2 -4 u^2 v^2} \pm (u^2-v^2). \label{SyEi}
\end{eqnarray}

From Eqs. (\ref{defIFT}) and (\ref{defGC}), for two Gaussian states $\rho $ and $\sigma $ with the zero means, their overlap can be expressed in terms of their covariance matrices as 
\begin{eqnarray}
\tr ( \rho  \sigma )=  \left[\det\left(\frac{\gamma_\rho+ \gamma_\sigma}{2}\right)\right]^{- 1/2}.   
\end{eqnarray}
From this relation, the expectation value of Eq. (\ref{GamG}) for a Gaussian state $\rho $ can be written as  
\begin{eqnarray}
\ave{\Gamma(\hat G_\lambda' )}= \tr {\rho \Gamma(\hat G_\lambda ') }
 &=& \left[\det\left(\frac{\gamma_\rho+ \gamma_0}{2}\right)\right]^{- 1/2} , 
\end{eqnarray}
and the overlap separable condition of Eq. (\ref{LGTP2})  turns out to be % \begin{eqnarray}F\le \frac{\lambda}{1+ \lambda}\end{eqnarray} leads to
\begin{eqnarray}
  \det\left(\frac{\gamma_\rho+ \gamma_0}{2}\right)&\ge &  1+\frac{2}{\lambda}+\frac{1}{\lambda^2}. \label{13siki} %\textrm{with }&& \lambda ^{-1} = \lambda ^{-1}%
% &&16\{\lambda^{-2} (n+m+2-2c_1)(n+m+2-2|c_2|) \nonumber \\
% &&+ \lambda^{-2}[(m+1)(n+1)- c_1^2][(m+1)(n+1)- |c_2|^2] \}
\end{eqnarray}
%\begin{eqnarray}\det\left(\frac{\gamma+ \gamma_0}{2}\right) \ge 1+2 \lambda ^{-1}+\lambda^{-2}\end{eqnarray}

The left-hand side of this expression can be simpler when  $\gamma_\rho$ is in the direct-sum form similar to the form of $\gamma_0$ in Eq. (\ref{bd1}). It is always possible to transform the covariance matrix into the direct-sum structure within the local Gaussian transformation \cite{Duan00,Simon00}.  
We thus consider the covariance matrix with the  following direct-sum form: \begin{eqnarray}
\gamma_\rho = \left(
  \begin{array}{cccc}
     n_1  &  0 &  c_1   & 0    \\
     0  & n_2   &  0 & c_2   \\
     c_1   &  0  &  m_1  & 0   \\
      0 & c_2   &  0  & m_2   \\
  \end{array}  \label{d-sum-form}
\right). \end{eqnarray}
Here, it is not necessary to consider an irreducible form with $n_1=n_2$ and $m_1=m_2$ as in \cite{Duan00,Simon00}. For the direct-sum form, the condition of Eq. (\ref{13siki}) leads to
\begin{eqnarray}
&& \det\left(\frac{\gamma_\rho+ \gamma_0}{2}\right) \nonumber\\ &=& \frac{1}{16 } \Big[ (n_1+1)(m_1 +1) - c_1^2  + \frac{2}{\lambda}( u^2 n_1 - 2uv c_1  +     v^2 m_1 )   \Big ]  \nonumber \\ &&\times \Big [ (n_2+1)(m_2+1)- c_2^2 +  \frac{2}{\lambda} (u^2 n_2 + 2 uv   c_2 + 
   v^2 m_2 )  \Big] %[(n+1+\frac{1}{\lambda})(m+1+\frac{1}{\lambda})- ( c_1+\frac{1}{\lambda} )^2]
  \nonumber \\
%&& \times [(n+1+\frac{1}{\lambda})(m+1+\frac{1}{\lambda})- (|c_2|+\frac{1}{\lambda} )^2] \nonumber \\ 
&\ge &  1+\frac{2}{\lambda}+\frac{1}{\lambda^2}. %\textrm{with }&& \lambda ^{-1} = \lambda ^{-1}%
% &&16\{\lambda^{-2} (n+m+2-2c_1)(n+m+2-2|c_2|) \nonumber \\
% &&+ \lambda^{-2}[(m+1)(n+1)- c_1^2][(m+1)(n+1)- |c_2|^2] \}
\end{eqnarray}
%\begin{eqnarray}(u^2 n  +v^2 m  -2 uv c_1)(u^2 n +v^2 m -2 uv |c_2|)  &\ge& 1 \nonumber \end{eqnarray}
 In the flat-distribution limit $\lambda \to 0$, %the coefficients of $\lambda ^{-2}$ are left and
  we obtain the following condition:
\begin{eqnarray}
 && \frac{1}{4}( u^2 n_1+  v ^2 m_1  -2c_1 u v +1 )\nonumber \\&\times& (  u^2 n_2+  v^2 m_2 +2 c_2  u v +1 ) \ge 1.
\end{eqnarray}
This condition is simply expressed in terms of the EPR-like uncertainties as % expression in terms of $U$ and $V$   we obtain a simple  condition 
\begin{eqnarray}
 \frac{1}{4}( U  +1 )(  V +1 ) \ge 1. \label{uv3}
\end{eqnarray}
where we use  Eq. (\ref{defUV}) with $C=1/2$, namely, we use
\begin{eqnarray}
U &=&  {2} \ave{\Delta^2  \left( u \hat x_A - v  \hat x_B \right) }= (u^2 n_1  +v^2 m_1  -2 uv c_1) \nonumber \\ 
V &=& 2 \ave{\Delta^2 \left( u  \hat p_A + v  \hat p_B \right)}= (u^2 n_2 +v^2 m_2 + 2 uv  c_2 ). \nonumber \\ \label{ssf}
\end{eqnarray}
Therefore, a bit surprisingly, it turns out that the overlap condition can be  seen as a separable condition with the EPR-like uncertainties.

Taking square root on both sides of Eq. (\ref{uv3}) and using the relation $U+V+2 \ge 2 \sqrt{U+1} \sqrt{V+1}$ we can reproduce Eq. (\ref{uv2}), i.e., we have $U+V \ge 2$ again. It implies that, for the class of Gaussian states expressed in the standard form, the overlap condition is tighter than the sum condition.  
 On the other hand, it is known \cite{Duan00} that any two-mode entangled Gaussian state can be detectable by the sum condition by using a specific standard form also written in the  form of Eq. (\ref{d-sum-form}). Therefore, it is concluded that the overlap condition is capable of detecting the inseparability of two-mode Gaussian states.

\begin{figure}[tbph]
\includegraphics[width= .8\linewidth]{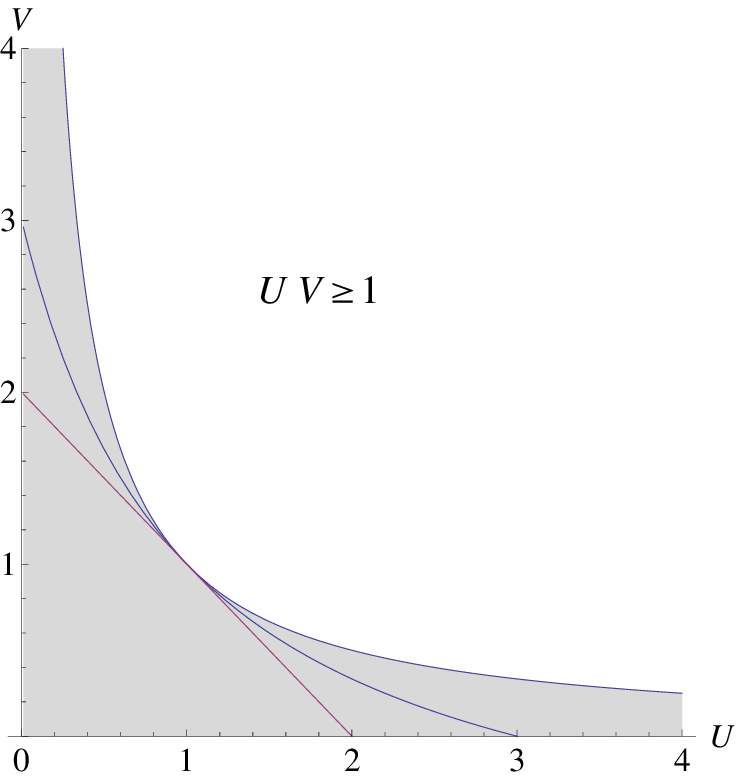} %,epp-ur-fig1.eps}%{epp-fig-c.eps}
  \caption{(Color online) The three separable conditions for the Gaussian states  are described by the three curves on the $U$-$V$ plane associated with the variances for the EPR-like operators $(U,V)$. The three curves are inscribed at the single point $(U,V) = (1,1)$. 
    The product condition of Eq. (\ref{prod-c}) implies that any state is entangled if its location on the $U$-$V$ plane $(U,V)$ is below the inverse proportional curve $UV = 1$ (gray regime). The boundary of the sum condition of Eq. (\ref{sum-c}) is given by     the straight line $U + V  = 2 $. The boundary of the overlap condition of Eq. (\ref{LGTP2}) for the Gaussian states in the standard form is given by $(U+1)(V+1)=4 $, which includes the three points $(1,1)$, $(3,0)$, and $(0,3)$.  This curve is lying in the middle of the two lines. It shows that the performance of the overlap condition is in between the product condition and the sum condition.  \label{fig:epp-figc21.eps}}
\end{figure}

We can also show that the product condition of Eq. (\ref{uv1}) can reproduce the overlap condition of Eq. (\ref{uv3}) as follows. From Eq. (\ref{uv1}) we have $UV + 2 \sqrt{UV} +1 \ge 4$. From this relation and $U+V \ge 2 \sqrt{UV}$ we have $UV + U+V+1 \ge  4$. This relation is nothing but Eq. (\ref{uv3}). We thus have proven the embrace relation for the three separable conditions: 
\begin{eqnarray}
\textrm{Eq.} \ (\ref{uv1}) \subset  \textrm{Eq.} \  (\ref{uv3}) \subset   \textrm{Eq.}\ (\ref{uv2}).  \label{EBR}
\end{eqnarray}
This embrace relation can be displayed on the $U$-$V$ plane in FIG. \ref{fig:epp-figc21.eps}. Thereby, geometrically we can prove that the overlap condition is tighter than the sum condition and that the product condition (\ref{prod-c}) is tighter than the overlap condition.

\section{Experimental measurement schemes} In this section we describe how to estimate the state-overlap to the EPR-like correlated classical mixture $\ave{\Gamma G_\lambda'}$ of Eq. (\ref{GamG}) %[ or Eq. (\ref{eq22})]
 in experiments.

The heterodyne measurement corresponds to a projection to coherent states and its positive-operator-valued-measure elements are symbolically written as $\{ \ketbra{\alpha}{ \alpha}/ \pi\}$. If we perform heterodyne measurement   each of the two modes $A$ and $B$, then we can obtain the joint probability distribution associated with the projection to the product of the coherent states $\ket{\alpha}_A\ket{\beta}_B$ where $\alpha$ and $\beta$ correspond to the outcomes of the measurement on the system $A$ and the system $B$, respectively.  
From this joint probability distribution,  the strength of the coherent-state-based EPR-like correlation $\ave{\Gamma \hat G_\lambda '}$ for any pair of the parameters $(u,v )$ can be calculated. This enables us to check the overlap separable condition of Eq. (\ref{LGTP2}) in principle. This may not be an efficient way for entanglement detection because the heterodyne statistics include the information of the full-tomographic reconstruction \cite{Leonhardt}.

\begin{figure}[htbp]
\includegraphics[width=1\linewidth]{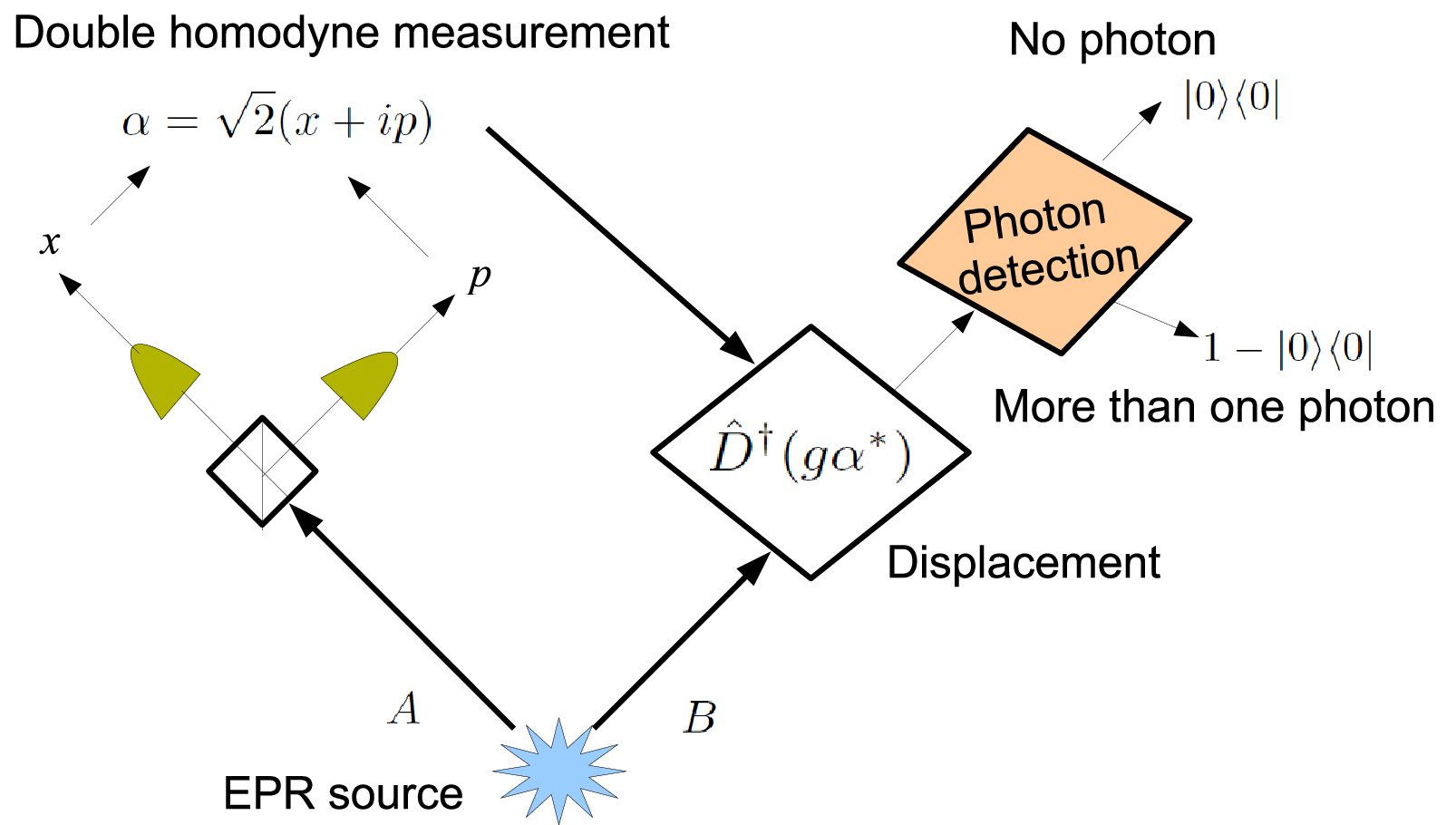}%{epp-fig3-1.eps}
  \caption{(Color online) The projection probability to a pair of coherent states $\ket{\alpha} \ket{g\alpha ^*}$ can be measured by a heterodyne (double homodyne) measurement and a photon detection. The measurement outcome of the heterodyne measurement $(x,p)$ determines the amplitude of the coherent state $\alpha = \sqrt 2 (x+ip )$ of the mode $A$.   The photon detection determines whether the number of the photon in the measured mode is zero or more than 1; It is the projection to one of the two subspaces  $\ketbra{0}{0}$ and $I - \ketbra{0}{0}$ where $I= \sum_{n=0}^{\infty } \ketbra{n}{n}$. The photon detection of the mode $B$ after the displacement operation $\hat D^\dagger (g \alpha ^*)$ gives the information whether or not the state is initially in the coherent state  $\ket{g \alpha ^*}= \hat D(g \alpha ^*)\ket{0}$ because this state becomes vacuum after the displacement as $\hat D ^\dagger (g \alpha ^*) \ket{g \alpha ^*}= \ket{0}$.         }% $ D (\alpha ^*)$ \ \ \   $1- \ketbra{0}{0}$ \ \ \   $\ketbra{0}{0}$ 
 \label{fig:epp-fig3.eps}
\end{figure}
 In turn, if a pair of the parameters $(u,v )$ is specified beforehand, we  only need to consider the probability associated with  the specific pairs of the states  $\ket{\alpha}_A\ket{g \alpha^* }_B$ with $g = v/u$. In order to measure this probability, a possible measurement process is composed of a heterodyne measurement and a photon detection with a feedforward control as in  FIG. \ref{fig:epp-fig3.eps}. We first perform the heterodyne measurement on the system $A$. Then, according to the outcome of the heterodyne measurement $\alpha$, we apply the displacement operation with an amount of the displacement $g \alpha ^*$ on the system $B$. Finally, we perform the photon detection of the system $B$.  
 It confirms whether or not the system $B$ was $\ket{g \alpha ^*}$. This is  because the displacement $\hat D^\dagger  (g \alpha ^* )$ transforms the coherent state $ \ket{g \alpha ^*}$ to the vacuum state as $\ket{0 }= \hat D^\dagger  (g \alpha ^* )\ket{g \alpha ^*}$ and the vacuum state is correctly discriminated by the photon detection as the no-photon event.
{ This measurement technique has been demonstrated recently  \cite{Wittmann08, Tsujino10,Tsujino11}.} Repeating this process we can obtain the probability that the pair state $\ket{\alpha}_A\ket{g \alpha^*}_B$ is contained in the total system initially. From the measured expectation values $\ave{\ketbra{\alpha }{\alpha } \otimes \ketbra{g \alpha^* }{g \alpha^*}}$, a class of the separable conditions associated with   Eq. (\ref{eq22}) can be checked as well.

\section{Conclusion}
We have introduced the notion of the coherent-state-based EPR-like correlation as a state-overlap to a classically correlated coherent-state mixture. We have shown that the separable condition with this state-overlap is capable of  detecting entanglement of two-mode Gaussian states. The separable threshold was derived by using the partial transposition on a limitation of the phase-space localization. A parallel formulation was given for the product separable condition concerning the standard EPR-like correlation. 
  We have also addressed how to detect the state-overlap experimentally by the heterodyne detection and the following photon detection with a feedforward control.

 %The author would like to thank   for helpful discussions.
%This work is partially supported by  .
This work was supported by GCOE Program ``The Next Generation of Physics, Spun from Universality and Emergence'' from MEXT of Japan. %the Grant-in-Aid for the Global COE Program ``The Next Generation of Physics, Spun from Universality and Emergence'' from the Ministry of Education, Culture, Sports, Science and Technology (MEXT) of Japan.%  supported by JSPS Research Fellowships for Young Scientists. 
 R.N. acknowledges support from JSPS.
%\end{theacknowledgments}

%ŽQl•¶Œ£


\begin{thebibliography}{99}
\bibitem{comS}The symplectic eigenvalues are obtained when set we set $\kappa =v/u$, $s= \lambda /u ^2$ on the expression of $\nu_\pm$ in Ref. \cite{namiki11R}.
\end{thebibliography}
\end{document}